# **Ways Towards Pure Deuterium Inertial Confinement Fusion Through the Attainment of Gigavolt Potentials**

F. Winterberg

*University of Nevada, Reno*

**Abstract**

The attainment of ultrahigh electric potentials by suppressing the stepped leader breakdown of a highly charged conductor levitated in a spiraling Taylor flow opens up the possibility of order of magnitude larger driver energies for the ignition of thermonuclear reactions by inertial confinement. In reaching gigavolt potentials, intense $10^{16}$ Watt, GeV ion beams become possible. Together with their large self-magnetic field, these beams should be powerful enough to launch a thermonuclear micro-detonation into pure deuterium, compressed and ignited by such beams. In high gain laser fusion the proton flash from the micro-explosion is likely to destroy the optical laser ignition apparatus, and it is not explained how to avoid this danger. The possible attainment of gigavolt potentials could make laser fusion obsolete.

## 1. Introduction

The main efforts in inertial confinement fusion research are at the present directed towards improved targets, rather than better drivers. The focus on targets is motivated by the still comparatively low energy of lasers as the most popular driver. Laser beams can be easily focused onto a small area. What stands against their good stand-off property, is that for a high gain the intense photon flashback from the ignited thermonuclear microexplosion can destroy the optical system needed for a repetition-rate operation. This problem adds to the poor laser efficiency. Generating a soft X-ray pulse by an electric pulse power driven wire array has a much higher efficiency, but a poor stand-off property. For that reason it is also not very suitable for a repetition-rate operation, because it would require replaceable transmission lines.

Back in 1968 I had proposed that ignition could be achieved by bombarding a small solid deuterium-tritium (DT) target with a 10 MJ – $10^{15}$ Watt ($10^{7}$ Volt, $10^{8}$ Ampere) relativistic electron beam drawn from a large Marx generator [1]. Because the current is there well above the Alfvén limit, the beam would have to be propagated through a current neutralizing background gas, making it more difficult to focus it onto the target. More serious was the problem how to dissipate the beam energy within the small target. Because of these difficulties, a radically different approach was proposed at the same time. It was to charge up to gigavolt potentials a magnetically levitated, and in ultrahigh vacuum magnetically insulated conductor of metersize dimensions, acting like a high voltage capacitor. Discharging this capacitor would make possible the generation of intense gigavolt ion beams, as the discharging of a Marx capacitor bank makes possible the generation of intense MeV electron beams.

Because of the considerable technical difficulties to sustain the required ultrahigh vacuum, a different approach to reach ultrahigh voltages would be highly desirable. It is suggested by the mechanism of natural lightning, where voltages up to $10^9$ Volt have been observed. Lightning can deliver an energy of several hundred megajoule, discharging several Coulombs with a current of 10 – 100 kA. It occurs if the electric field between a cloud and the ground exceeds the breakdown for air, ideally at about 30 kV/cm. For a 300 meter long lightning would imply a potential difference of $10^9$ Volt. Most lightning discharges are from a negatively charged cloud to the ground, but at rare occasions from the ground to a positively charged cloud. In these rare cases the current can reach 300 kA, discharging 300 Coulombs. At $10^9$ Volt, this is an energy of 300 gigajoule, equal to the energy released by 75 tons of TNT and with a power of $\cong 3\times 10^{14}$ Watt. By comparison, to ignite a thermonuclear micro-explosion in liquid deuterium-tritium requires an energy of about 10 MJ with a power of $10^{15}$ Watt. This raises the question if one cannot make artificial lightning, comparable in energy and power of natural lightning, to drive inertially confined thermonuclear micro-explosions. As mentioned, one way it conceivably can be done is by charging to gigavolt potentials a magnetically insulated conductor levitated in ultrahigh vacuum. This idea was elaborated in some further detail [2]. Here a quite different approach is proposed.

## 2. The Importance of High Voltages for Inertial Confinement Fusion

The reaching out for high voltages is of importance in the quest for the ignition of thermonuclear micro-explosions by inertial confinement for two reasons:

1. The energy e [erg] stored in a capacitor C [cm] charged to the voltage V [esu] is equal to

$$e = (1/2)\, CV^2, \tag{1}$$

with an energy density

$$\varepsilon \sim e/C^3 \sim V^2/C^2 \tag{2}$$

The e energy is discharged in the time τ [sec] (c velocity of light)

$$\tau \sim C/c, \tag{3}$$

with the power P [erg/s]

$$P \sim e/\tau \sim cV^2 \tag{4}$$

This shows that for a given dimension of the capacitor measured in its length, and hence volume, the energy stored and power released goes in proportion to the square of the voltage.

2. If the energy stored in the capacitor is released into a charged particle beam with the particles moving at the velocity v, the current should be below the critical Alfven limit

$$I = \beta\gamma I_A \tag{5}$$

where $\beta = v/c$, v particle velocity, $\gamma = (1-v^2/c^2)^{-1/2}$ the Lorentz boost factor, and $I_A = mc^3/e$. For electrons $I_A = 17$ kA, but for protons it is 31 MA. For $I \ll \beta\gamma I_A$, one can view it as a beam of particles accompanied by an electromagnetic pulse, while for $I \gg \beta\gamma I_A$ it is better viewed as an electromagnetic pulse carrying along with it some particles. For $I \gg \beta\gamma I_A$, the beam can propagate in a space-charge and current-neutralizing plasma, but only if $I \leq \beta\gamma I_A$ can the beam be easily focused onto a small area, needed to reach a high power flux density. If a power of ~ $10^{15}$ Watt shall be reached with a relativistic electron beam produced by a $10^7$ Volt Marx generator, the beam current would have to be $10^8$ Ampere with $\gamma \cong 20$ and $\beta\gamma I_A \sim 3\times 10^5$ Ampere, hence $I \gg \beta\gamma I_A$. But if the potential is $10^9$ Volt, a proton beam accelerated to this voltage and with a current of $I = 10^7$A is below the Alfven current limit for protons. It would have the power of $10^{16}$ Watt, sufficiently large to ignite a deuterium thermonuclear reaction.

## 3. Electric Breakdown below the Paschen Limit as an Electrostatic Instability

According to Paschen's law the breakdown voltage in gas between two plane parallel conductors is only a function of the product pd, where p is the gas pressure and d the distance between the conductors. For dry air and a pressure of 1 atmosphere, the breakdown voltage is $3\times 10^6$ V/m, such that for a pressure of 100 atmospheres, it would be $3\times10^{8}$ V/m. In reality the breakdown voltage is much smaller. The reason is that by a small initial inhomogeneity in the electric field, more negative charge is accumulated near the inhomogeneity further increasing the inhomogeneity, eventually forming to a "leader", a small luminous discharge of electrons bridging part of the distance between the electrodes with a large potential difference. As a result a much larger electric field inhomogeneity is created at the head of the "leader", which upon repetition of the same process forms a second "leader", followed by a third "leader", and so on, resulting in a breakdown between the electrodes by a "stepped leader". This happens even though the electric field strength is less than the field strength for breakdown by Paschen's law. What one has here is a growing electrostatic instability, triggered by a small initial electric field inhomogeneity. A preferred point for the beginning of a stepped leader is the field inhomogeneity near the triple point where the conductor, the gas and the insulator, meet.

## 4. Stabilization in a Drag-Free Taylor flow

It is known, and used in electric power interruptors that a high pressure gas jet under can blow out an electric arc discharge. Recognizing the breakdown below the Paschen limit as a growing electrostatic instability, it is conjectured that much higher voltages can be reached if the onset of this instability is suppressed by a gas flow, with the stagnation pressure of the flow exceeding the electric pressure in between the electrodes, thereby overwhelming the electric pressure of a developing electric field inhomogeneity. It is for this reason proposed to levitate a spherical conductor by both hydrodynamic and magnetic forces inside a Taylor flow [4], a special drag-free spiral flow (see Fig.1).

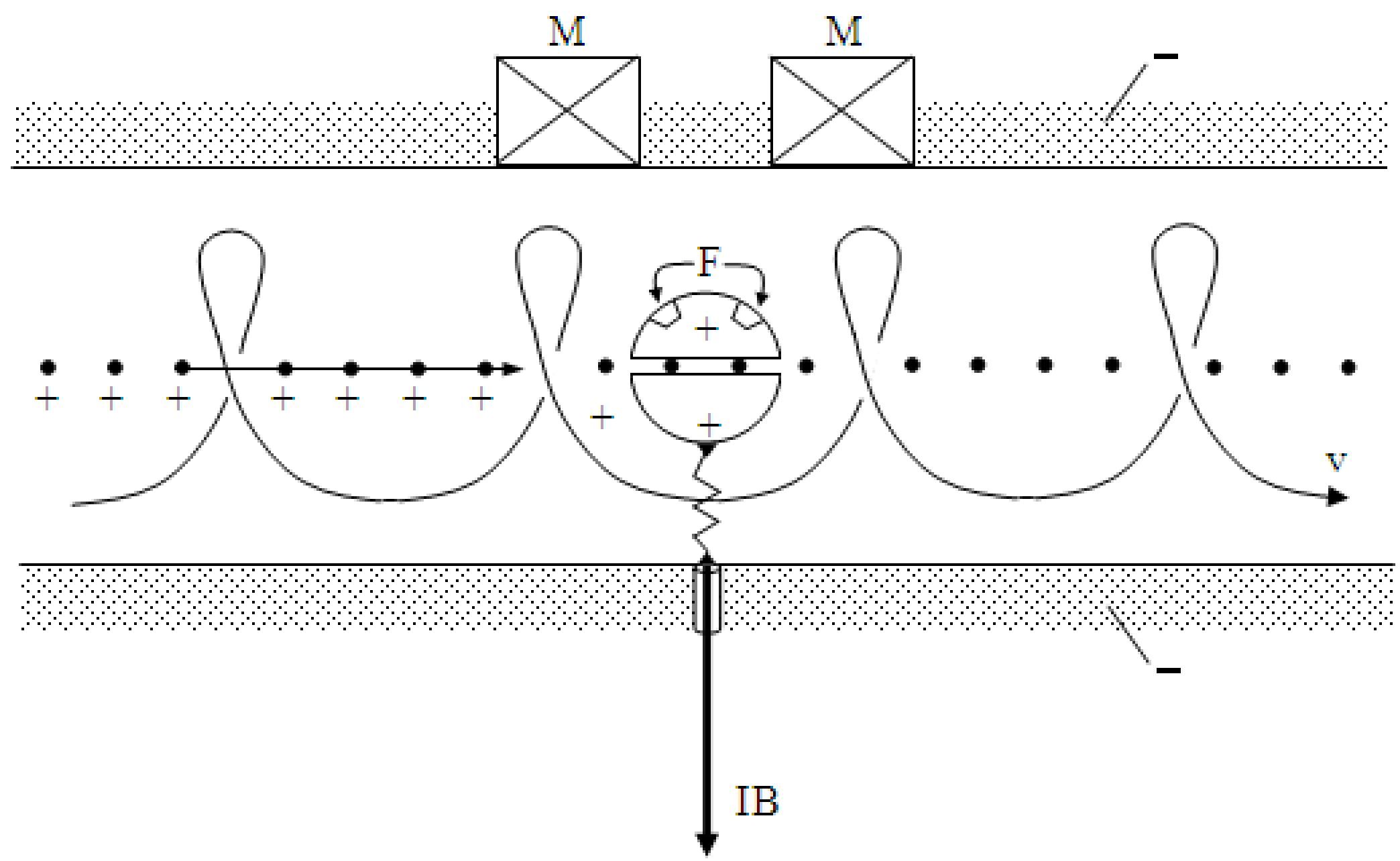

Fig.1. In the drag-free Taylor flow a magnetically levitated sphere is charged to ultrahigh potentials by electrically charged pellets passing through the center of the sphere, **M** magnets, **F** ferromagnets, **IB** ion beam.

In the absence of horizontally directed drag forces on the sphere inside the spiraling Taylor flow, the sphere has still to be levitated in the vertical direction by an externally applied magnetic field. By its levitation, the triple point as the source of a field inhomogeneity is eliminated.

For a pressure of 300 atmospheres the breakdown voltage would be $10^9$ V/m = $3\times10^4$ [esu],with an electric pressure $E^2/8\pi \sim 4\times10^7\, dyn/cm^2 \sim 40$ atmospheres. At a pressure of 300 atmospheres, gases at room temperature have a density of the order $\rho \sim 1 g/cm^3$. For the stagnation pressure $p = (1/2)\rho v^2$ of the Taylor flow moving with the velocity v [cm/s], to exceed the electric pressure, requires that $(1/2)\rho v^2 > E^2/8\pi$, from which one obtains for the given example that $v \geq 100$ meters per second.

The electric field strength at the surface of the meter-size conductor is $10^7$ V/cm, below $10^8$ V/cm where field ion emission sets in. For a ten times smaller velocity one could reach $10^8$ Volt, with the other parameters remaining the same. Instead of a gas under high pressure one may also use a nonconducting fluid under normal pressure.

What remains is how to charge the sphere to such a high electric potential. There seem to be two possibilities:

1. Similar as in a Van de Graaff generator, by letting a stream of positively charged pellets pass through of the sphere, releasing their charge in the center of the sphere.
2. By inductive charging in a rising magnetic field, releasing charges from the center of the sphere to the fluid flowing through the sphere in a duct [5] (see Fig. 1).

The highly charged sphere can be discharged over a spark gap, to be triggered by moving the sphere towards the wall with the help of the magnetic field holding the sphere in the center of the Taylor flow, until the pd product becomes smaller than the value where breakdown sets in below the Paschen curve. If the sphere is positively charged, and if the discharge current is larger than the Alfven current for electrons, (but smaller for ions), this will favor a discharge into an intense ion beam suitable as a driver for inertial confinement fusion.

**5. Pure Deuterium Inertial Confinement Fusion**

With no deuterium-tritium (DT) micro-explosions yet ignited, the ignition of pure deuterium (D) fusion explosions seems to be a tall order. An indirect way to reach this goal is by staging a smaller DT explosion with a larger D explosion in a mini Teller-Ulam configuration. A direct way for pure deuterium burn requires a driver with order of magnitude larger energies. It can be provided through the generation of GeV potentials.

Because the current needed for ignition is below the Alfven limit for ions, the beam is "stiff". The critical Alfven current for protons is well in excess of the critical current to entrap the DD fusion reaction products, a condition for detonation [6]. For a detonation in deuterium to occur the burn of the tritium and $He^3$ fusion products of the D-D reaction is important [7].

As shown in Fig. 1, a gigajoule intense relativistic ion beam below the Alfven current limit can be released from a to GeV potential charged conductor inside the Taylor flow, and directed onto the D explosive for its ignition. For a proton beam, the breakdown from the charged conductor must be in hydrogen, with the beam propagating towards the target in a space-charge neutralizing gas. Because the current needed for ignition is below the Alfven current limit for ions, the beam is "stiff".

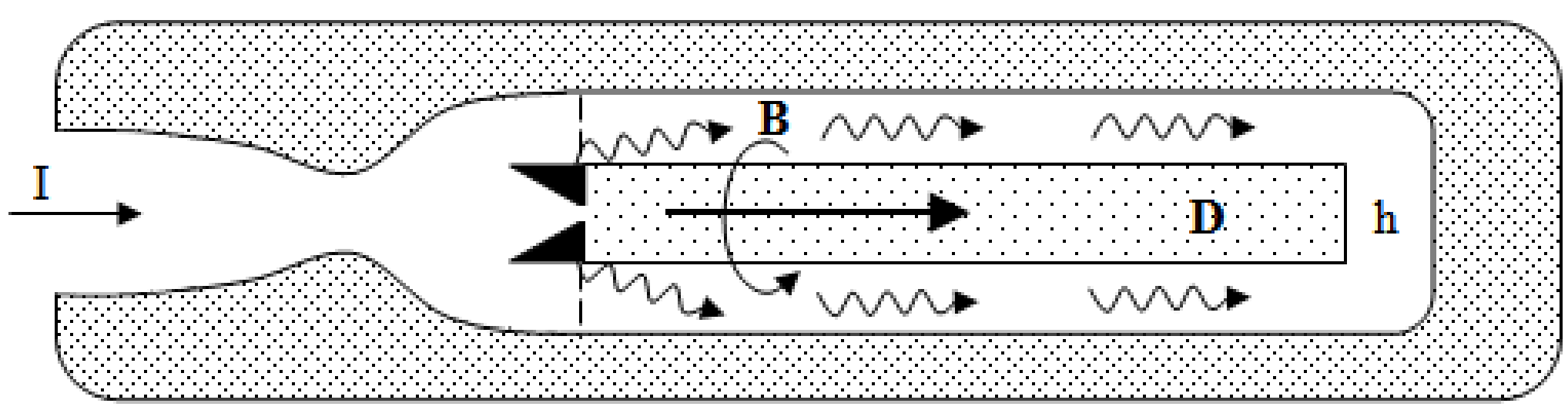

Fig.2: Pure deuterium fusion explosion ignited with an intense ion beam. **D** solid deuterium rod, **h** hohlraum, **I** ion beam, **B** magnetic field.

In a possible configuration shown in Fig.2, the liquid (or solid) D has the shape of a cylinder, placed inside a cylindrical "hohlraum" **h**. A GeV proton beam **I** coming from the left, in entering the hohlraum dissipates part of its energy into a burst of X-rays, compressing the deuterium cylinder, and part of it igniting a detonation wave propagating down the cylinder. With a gigajoule energy pulse lasting less than $10^{-7}$ s, the beam power is greater than $10^{16}$ Watt, sufficiently large to ignite a deuterium detonation along the cylinder.

If the rod has the length z and the density ρ, the ignition condition for deuterium requires that $\rho z > 10$ g/cm$^2$, and that the deuterium is heated over this length to a temperature T $\cong 10^{9\circ}$ K. Normally, the ρz condition is given by a ρr condition for the radius of a deuterium sphere. Here however, the radial entrapment of the charged D-D fusion reaction products, ensured by the magnetic field of the proton beam current in excess of $10^7$ Ampere, reduces the $\rho r > 10$g/cm$^2$ condition, to a $\rho z > 10$g/cm$^2$ condition, which is much easier to achieve. The yield of the deuterium explosion then only depends on the total mass of the deuterium.[1]

With both the beam and the target at a (initially) low temperature, the stopping length is determined by the electrostatic two stream instability [8], which in the presence of the strong azimuthal magnetic field set up by the proton current, it is enhanced by the formation of a collisionless shock [9]. The stopping range of the protons by the two stream instability is given by

$$\lambda \cong \frac{1.4c}{\varepsilon^{1/3}\omega_i} \tag{6}$$

where c is here is the velocity of light, $\omega_i$ the proton ion plasma frequency, and $\varepsilon = n_b/n$, where $n_b$ is the proton number density in the proton beam, and n the deuterium target number density. If the cross section of the beam is 0.1 cm$^2$, one obtains $n_b = 2\times10^{16}cm^{-3}$. For a 100 fold compressed deuterium rod one has $n = 5\times10^{24}cm^{-3}$ with $\omega_i = 2\times10^{15}s^{-1}$. One there finds that $\varepsilon = 4\times10^{-9}$ and $\lambda \cong 1.2\times10^{-2}cm$. At a deuterium number density $n = 5\times10^{24}cm^{-3}$, the deuterium density is $\rho = 17 g/cm^3$. To have $\rho z \geq 10 g/cm^2$, thus requires that $z \geq 0.6cm$. With $\lambda < z$, the condition for the ignition of a thermonuclear detonation wave is satisfied.

[1]For the DT reaction one must have $\rho\gamma \geq 1$g/cm$^2$ and T> 10keV.

The ignition energy is given by

$$E_{ign} \sim 3nkT\pi r^2 z \tag{7}$$

For 100 fold compressed deuterium, one has $\pi r^2 = 10^{-2} cm^2$, when initially it was $\pi r^2 = 10^{-1} cm^2$.

With $\pi r^2 = 10^{-2} cm^2, z = 0.6cm, kT \cong 10^{-7} erg$ (T~$10^9$°K), one finds that $E_{ign} \cong 10^{16} erg = 1GJ$. This energy can be provided by a $10^7$ Ampere $10^9$ Volt proton beam lasting $10^{-7}$ seconds. The power of $10^{16}$ Watt is large enough for the ignition of the deuterium reaction, and the time short enough to assure the cold compression of deuterium to high densities. For a $10^3$ fold compression, found feasible in laser fusion experiments, the ignition energy is reduced to 100 MJ [10].

## 6. Taylor Flow

A Taylor flow is obtained by a superposition of an axial flow with the velocity U along the z-axis of a cylindrical r, z, φ reference system, and a constant swirl W= (U/$l$)r in the azimuthal direction, where $r/l$ is a measure of the strength of the swirl. With the stream function $\psi$ ($r, z$), satisfying the equation [10]

$$\frac{\partial^2\psi}{\partial z^2} + \frac{\partial^2\psi}{\partial r^2} - \frac{1}{r}\frac{\partial\psi}{\partial r} + \frac{4}{l^2}\psi = \frac{2U}{l^2}r^2 \tag{8}$$

the velocity components in the z, r and φ- direction are

$$\left.\begin{aligned}\frac{u}{U}&=\frac{1}{r}\frac{\partial\psi}{\partial r}\\ \frac{v}{U}&=\frac{1}{r}\frac{\partial\psi}{\partial z}\\ \frac{w}{U}&=\frac{2}{lr}\psi\end{aligned}\right\}\qquad(9)$$

For a different problem the solution of (8), has been given in terms of Bessel and Neumann functions, by Moore and Leibovich where are $\kappa_1$, $\kappa_2$ constants of integration [11]

$$\psi=\frac{1}{2}Ur^2\left(1+\kappa_1\frac{J_{3/2}(\xi)}{\xi^{3/2}}+\kappa_2\frac{N_{3/2}(\xi)}{\xi^{3/2}}\right),\ \xi=\frac{2}{l}\sqrt{r^2+z^2}\qquad(10)$$

$$J_{3/2}(\xi)=\sqrt{\frac{2}{\pi\xi}}\left(\frac{1}{\xi}\sin\xi-\cos\xi\right),\ N_{3/2}(\xi)=\sqrt{\frac{2}{\pi\xi}}\left(\sin\xi+\frac{1}{\xi}\cos\xi\right).$$

The velocity components in the z, r and φ direction are then

$$\frac{u}{U}=1+\kappa_1\left(\frac{J_{3/2}}{\xi^{3/2}}-2\frac{r^2}{l^2}\frac{J_{5/2}}{\xi^{5/2}}\right)+\ldots\qquad(11)$$

$$\frac{v}{U}=2\kappa_1\frac{zr}{l^2}\frac{J_{5/2}}{\xi^{5/2}}+\ldots\qquad(12)$$

$$\frac{w}{U}=\frac{r}{l}\left(1+\kappa_1\frac{J_{3/2}}{\xi^{3/2}}\right)+\ldots\qquad(13)$$

the terms involving the Neumann functions are represented by dots. The J-solution of (10), putting $\kappa_2 = 0$, has no singularity at $\xi = 0$, and is the solution of interest. With the appropriate boundary condition it is:

$$\psi=\frac{1}{2}Ur^2\left[1-\left(\frac{2R/l}{\xi}\right)^{3/2}\frac{J_{3/2}(\xi)}{J_{3/2}(2R/l)}\right]\qquad(14)$$

Now, if $J_{5/2}$ vanishes on the surface of a sphere placed in the Taylor flow, then all velocity components vanish on the surface of the sphere. And if $J_{5/2}$ is zero on the surface of the sphere, the circumferential shear vanishes as well. As a consequence, there is no boundary layer on the surface of the sphere and no drag. The pressure on the sphere is constant and the sphere stays at rest, except that it still is subject to a downward directed gravitational force, assuming U is directed horizontally. The downward force can be compensated by an externally applied magnetic field. This can be done by making part of the sphere from a ferromagnetic material.

## 7. Discussion

The idea to suppress high voltage breakdown through a stepped leader, by levitating a conductor to be charged up to a high electric potential in a drag- free Taylor flow, can be easily tested at

The experiment by Harvey [12], shown in Fig. 3, verifies the Taylor solution.

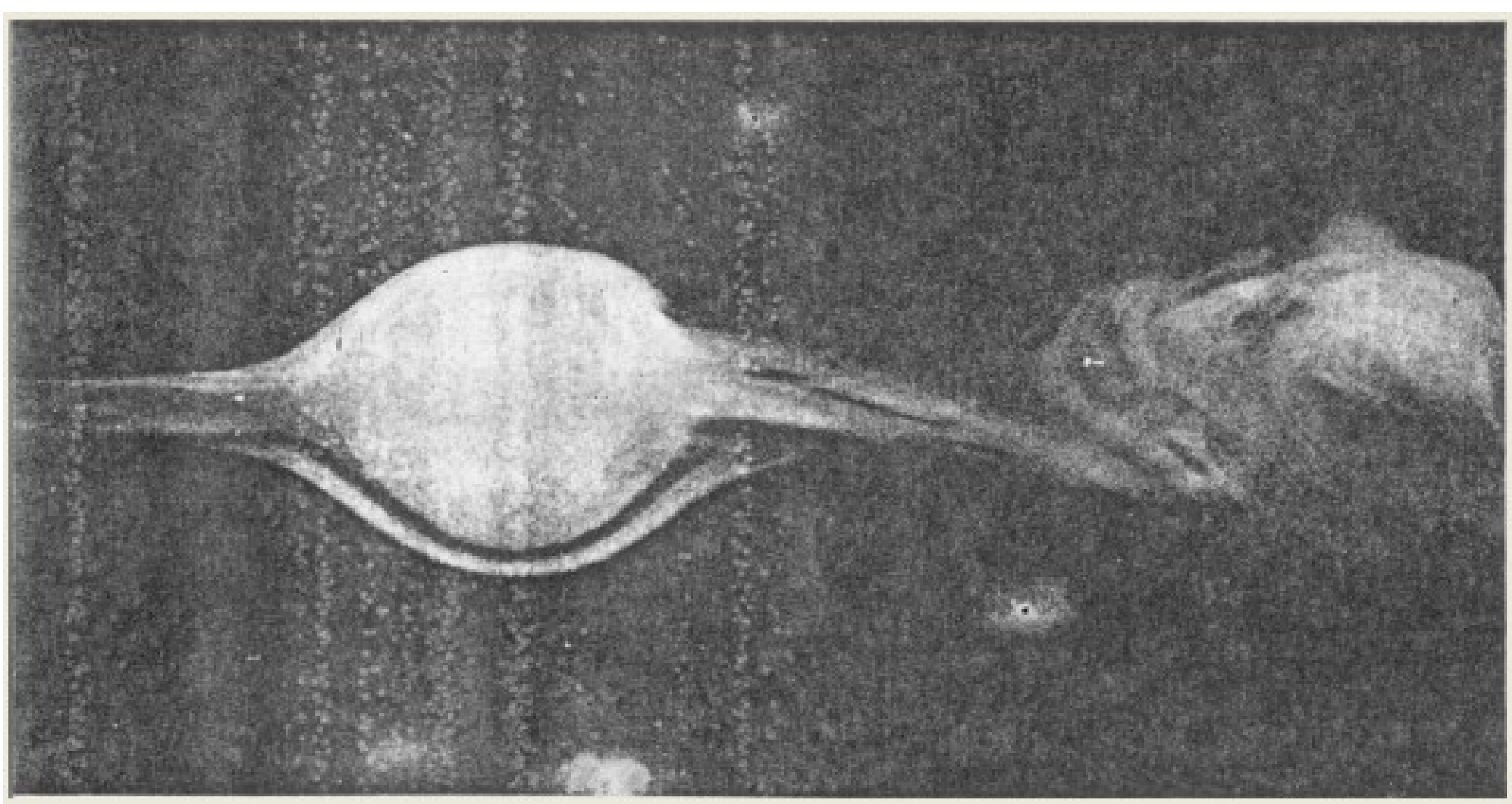

Fig.3 Experimental verification of the Taylor flow enclosing a spherical non-moving part inside the flow.

much lower voltages and flow velocities in the laboratory. And if it should turn out that the proposed concept will work, one can go to higher voltages, larger flow velocities, higher pressures and larger dimensions of the spherical conductor. One may also study what happens if the gas under high pressure is replaced by a liquid under normal pressure.

Snapshots of lightning discharges in the vicinity of tornadoes indicate that the lightning does not occur very near the tornado funnel, where the cloud to earth potential is greatest, but where also the air velocity is largest. There must be therefore an optimal distance for the breakdown to occur, near a location where the electric field, but not the air velocity, is still quite large.

## 8. Conclusion

If the proposed idea to generate ultrahigh voltages should turn out to be feasible, its importance for nuclear fusion can hardly be underestimated. With the prospect for non-fission ignited pure deuterium nuclear fusion micro-explosions, it could also become of interest for "plowshare" applications (peaceful uses of nuclear explosives).

GeV intense ion beams below the Alfvén limit would have a very good stand-off property for nuclear micro-explosions, and these beams would not suffer from the problem of laser beams, where in spite of their good stand-off property, the intense photon burst from a high gain micro-explosion can destroy the laser. The central problem for the achievement of the release of thermonuclear energy by inertial confinement is the energy and power of the driver. It can be easily met by a multi-gigajoule fission explosion, but should still be easy with a 10 to 100 MJ – 1GJ driver, with a power of $10^{16}$ Watt.